\documentclass[12pt,preprint]{aastex}

\def\ltsima{$\; \buildrel < \over \sim \;$}
\def\gtsima{$\; \buildrel > \over \sim \;$}
\def\lsim{\lower.5ex\hbox{\ltsima}}
\def\gsim{\lower.5ex\hbox{\gtsima}}
\def\lapp{\ifmmode\stackrel{<}{_{\sim}}\else$\stackrel{<}{_{\sim}}$\fi}
\def\gapp{\ifmmode\stackrel{>}{_{\sim}}\else$\stackrel{<}{_{\sim}}$\fi}

\newdimen\minuswidth    
\setbox0=\hbox{$-$}
\minuswidth=\wd0
\catcode`@=\active
\def@{\kern\minuswidth}
\setbox0=\hbox{\rm0}
 
\shorttitle{GeMS/GSAOI photometric and astrometric performance} 
\shortauthors{Dalessandro et al.}
 
\begin{document} 
\title{GeMS/GSAOI photometric and astrometric performance in dense stellar fields}

\author{
E. Dalessandro\altaffilmark{1,2},
S. Saracino\altaffilmark{2,1},
L. Origlia\altaffilmark{1},
E. Marchetti\altaffilmark{3},
F. R. Ferraro\altaffilmark{2},
B. Lanzoni\altaffilmark{2},
D. Geisler\altaffilmark{4},
R.~E Cohen\altaffilmark{4},
F. Mauro\altaffilmark{4,5},
S. Villanova\altaffilmark{4}
}
\affil{\altaffilmark{1} INAF --- Osservatorio Astronomico di Bologna, via Ranzani 1, I-40127, Bologna, Italy}
\affil{\altaffilmark{2} Dipartimento di Fisica \& Astronomia, Universit\`a degli Studi
di Bologna, viale Berti Pichat 6/2, I--40127 Bologna, Italy}
\affil{\altaffilmark{3} ESO - European Southern Observatory, 
             Karl Schwarzschild Str. 2, D-85748 Garching bei Muenchen, Germany}	  
\affil{\altaffilmark{4} Departamento de Astronom\'ia, Universidad de
  Concepci\'on, Casilla 160-C, Concepci\'on, Chile}
\affil{\altaffilmark{5} Millennium Institute of Astrophysics, Chile}

\begin{abstract}
Ground-based imagers at 8m class telescopes assisted by Multi conjugate Adaptive Optics 
are primary facilities to obtain accurate photometry and proper motions in dense stellar fields.
We observed the central region of the globular clusters Liller~1 and NGC~6624 
with the Gemini Multi-conjugate adaptive optics System (GeMS) feeding the  
Gemini South Adaptive Optics Imager (GSAOI) currently available at the Gemini South telescope, 
under different observing conditions. We characterized the stellar Point Spread Function (PSF)
in terms of Full Width at Half Maximum (FWHM), Strehl Ratio (SR) and Encircled Energy (EE), over the field of view.
We found that, for sub-arcsec seeing at the observed airmass, 
 diffraction limit PSF FWHM ($\approx$ 80 mas), SR $\sim40\%$ and EE $\ge50\%$ with a dispersion around $10\%$ 
over the 85\arcsec$\times$85\arcsec~field of view, can be obtained in the $K_s$ band. In the $J$ band the best images provide FWHMs between 
60 and 80 mas, SR $>10\%$ and EE $>40\%$. 
For seeing at the observed airmass exceeding 1\arcsec, the performance worsen but it is still possible to
perform PSF fitting photometry with $25\%$ EE in $J$ and $40\%$ in $K_s$.
We also computed the geometric distortions of GeMS/GSAOI and we obtained corrected images with an astrometric accuracy of 
$\sim$1 mas in a stellar field with high crowding.  
\end{abstract}

\keywords{Instrumentation: adaptive optics -- Instrumentation: high angular resolution -- Techniques: photometric --
	     Stars: imaging --  astrometry -- telescopes}


\section{Introduction}

Adaptive optics (AO) systems sample and correct in real time the wavefront deformation 
due to  the atmospheric turbulence, which affects the overall sharpness and spatial
resolution of the astronomical images obtained with ground-based telescopes. 

The first AO systems
were based on a single guide star, either the target itself if bright enough, or a bright star  
within a few arcsec from the astronomical target.
This system called single-conjugate AO (SCAO) 
however, only partially corrects the atmospheric turbulence because of the anisoplanatism, 
with best correction on axis and a blurring size of the astronomical image that 
increases with increasing distance from the guide star. 

In order to increase the field of view with good AO corrections, it is necessary to use   
multiple guide stars and correct for multiple layers of turbulence in the atmosphere.
Ground-based near IR (NIR) imagers assisted by multi-conjugate AO (MCAO) systems represent
the technological frontier of the last decade to obtain high quality stellar photometry in crowded fields 
at the highest possible spatial resolution, by reaching the diffraction limit of the 8m class telescopes.

The advantage of using a MCAO with respect to a SCAO 
is significant: while normally the latter provides a  Strehl ratio (SR) above 20\%
within $20\arcsec$ from the guide star, with a MCAO system one can obtain such SRs over 
a field of view (FOV) as large as a few arcmin (\citealt{beck88}; \citealt{eller94}; \citealt{john94}; 
\citealt{lel02}; \citealt{mar07}).
  
The first MCAO system used for nighttime astronomical observations was the Multi-conjugate Adaptive-optics 
Demonstrator (MAD) operating at the VLT in 2007-2008 \citep{mar07}.
MAD used up to three natural guide stars for the wavefront sensing and tip-tilt correction and two deformable
mirrors conjugated at the ground and at an altitude of 8.5~Km, providing a corrected FOV of about
$1\arcmin \times 1\arcmin$. 
A few works on stellar photometry in the dense stellar fields of Galactic globular clusters 
\citep[see, e.g.][]{ort11,fer09,mor09,bono10} have shown 
how effective a MCAO system like MAD could  be in providing uniform Point Spread Functions (PSF) 
and accurate photometry across the entire $1\arcmin \times 1\arcmin$ FOV.

Since 2013 the Gemini Multi-conjugate adaptive optics System (GeMS) together with the 
Gemini South Adaptive Optics Imager (GSAOI) at the Gemini South telescope 
\citep{rig12,rig14,nei14a,nei14b} is regularly offered to the Community for observations. 
This is the only MCAO facility currently at work in the world.
GeMS is the first sodium based multi-laser MCAO system. It uses five lasers and three tip-tilt stars 
to provide an efficient correction over a $\sim 1.5\arcmin\times1.5\arcmin$ FOV.
GSAOI is a NIR imager equipped with four $2k\times2k$ detectors 
with 20 mas pixel size, 
covering $85\arcsec \times 85\arcsec$, designed to work at the diffraction
limit of an 8m telescope.

The proper characterization of the image quality delivered by the GeMS/GSAOI system and of the parameters 
that mostly contribute to set its overall efficiency 
is extremely important to decide the best observational strategy and to maximize the scientific output. 
It also provides useful information for  the future generation of MCAO systems at the 20-40m class giant 
telescopes currently underway.

Indeed various authors have put significant effort on this task.
In particular Neichel et al. (2014a; see also \citealt{rig12,rig14} and \citealt{vidal13}) have analyzed 
the average performance of GeMS in terms of
SR and Full Width at Half Maximum (FWHM) variations for a large sample of images acquired during the
Science Verification. They find that, with a median seeing of $0.73\arcsec$,  the average FWHM delivered by the
system for the 50\% of the images is 0.087\arcsec,
0.075\arcsec and 0.095\arcsec in the $J$, $H$ and $K_s$ bands respectively. For reference, the diffraction limited
FWHM are  e.g. 0.037\arcsec, 0.049\arcsec and 0.068\arcsec at 1.2, 1.6 and 2.2 $\mu$m, respectively. 
They also find that the average
FWHM variation over a field of one square arcmin is  $\sim 5\%$ (relative r.m.s.) 
with the maximum variation being  $\sim15\%$. For the same images the average SR is $5\%$
in $J$ and  $17\%$  in $K_s$. 
More generally, the SR and FWHM can vary by a factor of 2-3 (also depending on the considered filter)
for seeing variation in the range 0.5\arcsec - 1.5\arcsec.

In addition to the natural seeing, there are other physical parameters that
can affect the performance of the GeMS/GSAOI and AO systems in general.
Among them, a non-negligible role is played by the natural guide stars (NGS)  brightness and asterism, the 
laser guide star (LGS) photon return, turbulence speed and profile. 
In particular, \citet{vidal13} and \citet{nei14a} have shown that the seasonal change of the LGS photon return can affect
the delivered average FHWM and SR values by up to a factor of 2-3. 
On the same line, \citet{vidal13} illustrate a case where for the same targets, same photon return and natural seeing, 
the SR drops by a factor of two, most likely due to variations of the atmospheric turbulence profile ($C_N^2$).

These works clearly demonstrate the importance and complexity of disentangling the impact of different factors 
on the final performance of the AO system. 
By following on these first characterizations, we present here a complementary analysis with the aim of looking at the
GeMS/GSAOI performance from an observer/user perspective and providing additional information to be eventually used 
for Phase I and Phase II preparation. 
To this aim, we use a sample of images acquired within a scientific proposal devoted to  
the study of the properties and stellar content of a sample of Galactic bulge globular clusters. 
First results from this project have been recently published by \citet{sar15,sar16}.

In addition to the FWHM and SR average values that have been analyzed also in previous papers, 
we examine the behavior of the EE as a function of natural seeing. This quantity is a very intuitive 
parameter characterizing
the properties of the PSF and it can be directly compared to diffraction-limited space telescopes as well as 
ground-based instruments not supported by AO facilities. 
We also add a systematic analysis about how natural seeing impacts the uniformity of the PSF.
For the first time we tentatively account for the role played by the airmass and NGS brightness and we present the first analytic solution for the 
geometric distortions of the GeMS/GSAOI system.

The paper is structured as follows: in Section~2 we describe the data-set and data-analysis;
in Section~3 we provide a characterization of the GeMS/GSAOI performance by using 
the PSF FWHM, SR and EE as figures of merit and a comparison with Hubble Space Telescope (HST) images; 
in Section~4 we analyze the geometric distortions and in Section~5 
we draw our conclusions.

\section{Data-set and observing conditions}

By using GeMS/GSAOI we observed the central regions of two Galactic bulge globular clusters Liller~1 and NGC~6624,
between April 2013 and May 2013 (Program ID: \dataset{GS-2013A-Q-23}; PI: D. Geisler). 
Two different sets of $J$ and $K_s$ images for Liller~1 and one set for NGC~6624 (see Table~\ref{tab1}) have been acquired
using an exposure time of 30 sec for each individual acquisition. 

Each image has been sky-subtracted and flat-field corrected by using suitable master sky and dome flat frames 
in the $J$ and $K_s$ filters.
We recall that each image is actually the mosaic of four chips that have been reduced and calibrated independently.

The atmospheric seeing at the zenith and at a given wavelength $\lambda$ can be computed using the formula:
$s(\lambda) = \lambda / R_0(\lambda) $.
We used the R$_0$ Fried parameter at $\lambda=500nm$ and at the zenith reported 
in each image header to obtain
$s(500nm) = 10.31 /R_0(500nm)$, 
where $R_0$ is in units of cm and the seeing in units of arcsec.

However, it is eventually useful to compute the seeing at the sky position of the target 
(i.e. at the observing airmass). 
We thus used the following formula
$s(\lambda,z) = \lambda / (R_0(\lambda) \times sec z^{-3/5}) $
to obtain
$s(500nm,z) = 10.31 /(R_0(500nm) \times sec z^{-3/5})$.
Seeing can be computed at other reference wavelengths by using the scaling relation
$s(\lambda)/s(500nm)=(\lambda/500)^{-0.2}$.
Hereafter, we always refer to seeing at 500nm.

It turns out that our images have been acquired under significantly different atmospheric conditions, 
with seeing $s(500nm)$ varying from
$\sim0.5\arcsec$ to $\sim1.5\arcsec$. 
In particular, the data of NGC~6624 have been obtained with an average
value of $s(500nm)\sim 0.65\arcsec$, while those for Liller~1 have been obtained under worse conditions 
with $s(500nm)\sim 1.00\arcsec$.

Three reference guide stars in each cluster (see Table~\ref{tab2}) for the tip-tilt correction have been selected.
The guide stars of NGC~6624 are likely luminous and cool cluster member giants near the tip of the RGB
with fmag$\sim$R = 13-14. 
The guide stars in Liller~1 are likely blue, foreground stars, 
with significantly fainter $J$, $H$, $K_s$ magnitudes and similar fmag$\sim$R band magnitudes compared to those of NGC~6624, 
with the exception of NGS~1, which is a couple of magnitudes fainter.
Since the R band is the spectral range where the wavefront is mostly sensed, this may have some relevance in the final  
performance (see Section~\ref{ave}).

To quantify how the LGS photon return changed during the observing nights, 
we use the LGS wavefront sensor photon counts reported in the header of each image.
First, it is important to stress here that all the images analyzed in this paper have been acquired 
mostly over two nights separated by a relatively short time interval 
(see Table~\ref{tab1}), while the LGS photon return is expected to vary mostly
seasonally (\citealt{vidal13}; \citealt{nei14a}). Also, about half of the images of Liller 1 were obtained in the same 
night as those of NGC 6624 (May 24 2013). 
As expected, we find a negligible variation of the average photon counts of the LGS during the observing nights.

\section{Overall performance}

We used the IDL-based \texttt{Multi-Strehl Meter} software written by E. Marchetti \citep{mar06}
to analyze the PSF of the science images and measure their FWHM, SR and EE with variable observing conditions,  
in order to characterize the performance of the AO system over a range of conditions. 

The first step of the analysis is the identification of the candidate star peaks. 
The software requires 
as input parameters a first-guess FWHM,  
the detection threshold and the size of the sub-image to search for the star peaks and to compute the local (residual) 
background.
The first-guess FWHM has been determined by computing the average FWHM
of a number of reference stars manually selected on each image. A detection threshold 
of 1000 ADU and a sub-image size of 60 pixels have been used.
Moreover, the software requires a few additional input parameters, among them the reference wavelength 
(1.25 $\mu$m and 2.15 $\mu$m for the $J$ and $K_s$ filters, respectively), the pixel size (20 mas), the telescope aperture (8.1m) 
and the obstruction factor of the primary mirror (12.35\%).

The obtained list of candidate stellar peaks for each image, 
suitably filtered by spurious detections, has been then cross-correlated 
with the photometric catalogs containing calibrated magnitudes and astrometric positions published by \citet{sar15} and
\citet{sar16} for Liller~1 and NGC~6624 respectively, in order to deliver 
the final list of stars to be analyzed.
Typically about 200 stars homogeneously distributed in the FOV in each image have been selected and their 
FWHM, SR and EE have been measured. 
The EE has been computed within a circular aperture of two times the measured FWHM,  
i.e. the typical aperture adopted in the photometric analysis\footnote{The Multi-Strehl Meter software actually yields 
the ensquared energy within 
2 times the FWHM. We then rescaled such a quantity to a circular aperture having the corresponding diameter.}.

We then computed average and corresponding dispersion values of the FWHM, SR and EE in each 
observed image and we analyzed their trend as a function of the seeing at the zenith 
($s(500nm$), see Figures~\ref{paramz} and ~\ref{errz}) and 
at the observing airmass ($s(500nm,z$), see Figures~\ref{parama} and ~\ref{erra}).

\subsection{PSF average properties}
\label{ave}

As shown in the left panels of Figure~\ref{paramz}, in the $K_s$ band average FWHMs 
very close to the diffraction limit of 70 mas, SR of $\sim40\%$ and EE of $55\%$ 
have been measured in the NGC~6624 images with sub-arcsec seeing and airmass close to one.
In the Liller~1 images with seeing between 0.9\arcsec and 1.5\arcsec, the average FWHM, SR and EE show 
a clear trend with the seeing but also quite a large scatter at a given seeing.
For seeing at the observing airmass increasing from $0.9\arcsec$ to $1.5\arcsec$, FWHM increases from 85 to 140 mas, 
while SR drops from 30\% to 12\% and EE from $50\%$ to $40\%$. 
The general observed trend, as well as the total range of values derived for NGC~6624 and Liller~1 in terms of both FWHM and
SR, is consistent with what found by \citet{vidal13} and \citet{nei14a} for $K_s$ images.
It is interesting to note that the largest average FWHM and 
the lowest SR and EE for a given seeing are measured in those images acquired at the largest airmass corresponding to $\sim1.4$. 
Hence, as shown in the left panels of Figure~\ref{parama}, 
when the FWHM, SR and EE are plotted against the seeing at the observing airmass, the scatter is reduced.
These measurements and comparisons indicate that also the airmass has an impact
on the delivered performance, and in this respect, it is worth noting that 
different observing airmasses could indeed explain some of the scatter observed in
Figure~6 by \citet{nei14a}, where their measured FWHMs and SRs are plotted against 
seeing at the zenith.

Most of the $J$ band images have been acquired at airmass close to one, hence 
their seeing at the observing airmass is very similar to the seeing at the zenith.
As shown in the right panels of Figures~\ref{paramz} and ~\ref{parama} the average FWHMs always 
exceed the diffraction limit of 40 mas, even with good seeing conditions of 0.6\arcsec and 
increases almost linearly with increasing the seeing.
The corresponding average SR and EE values decrease with increasing seeing.
For seeing increasing from $0.6\arcsec$ to $1.2\arcsec$, FWHM increases from 70 mas to 120 mas, 
while SR drops from $15\%$ to a few percent and EE from $40\%$ to $25\%$.
{Our FWHM and SR values indicate somewhat better performance of GeMS/GSAOI in the $J$ band with the respect to the findings by 
\citet{vidal13} and \citet{nei14a}, likely because of the uniform and good atmospheric conditions during our observations.

A sub-sample of images in the $J$ band of Liller~1 and NGC~6624 have been acquired with 
the same seeing, between $0.6\arcsec$ and $0.9\arcsec$ and can be used to check the impact of the different asterisms  
of the two clusters, and in particular the fact that one guide star in Liller~1 
has a significantly fainter R band magnitude (see Table~\ref{tab2}).
On average, the Liller~1 images show a $\sim10-15\%$ larger FWHM and smaller SR values,
that could be indeed a consequence of the significantly fainter guide star. 

Finally, we note that both in the $K_s$ and $J$ bands, 
the average EE values show a smoother variation with the seeing when compared to the variation of the FWHM and SR parameters.
This is somewhat expected, given that the EE is computed within a variable aperture, proportional to the variable FWHM, 
and it indicates that the seeing primarily impacts the spatial resolution (i.e. the FWHM and the SR) and to a lower extent 
the photometric signal (i.e. the EE), when computed via variable PSF fitting.

\subsection{PSF uniformity}  

The average values of the FWHM, SR and EE provide a measurement of the system efficiency, 
while their dispersion and spatial variation provides an estimate of the uniformity of the PSF across the FOV.
Modeling the PSF variations within the FOV is one of the major issues in the photometric analysis of crowded stellar fields in general, and 
especially when observed with ground-based AO-assisted imagers. Typical values of FWHM and SR variations can be found in the literature, 
in the following we quantify how their amplitudes vary as a function of the observing conditions.

As shown in Figures~\ref{errz} and \ref{erra}, 
the dispersion around the average FWHM increases with increasing seeing in a similar fashion as the FWHM itself.
This indicates that bad seeing worsens both the spatial resolution and its uniformity over the FOV. 
At variance, the dispersion around the average SR and EE decreases with increasing the seeing.
A more uniform SR and EE across the FOV with worsening seeing conditions is somewhat expected.
Indeed, at variance with the FWHM, the SR and the EE are quantities somewhat normalized to the seeing contribution.
Hence, when the seeing worsens, its contribution progressively dominates over the diffraction limit peak, 
and being practically constant across the FOV,
provides a progressively more uniform PSF.

In order to better visualize the spatial variations of the FWHM, SR and EE, 
in Figures~\ref{mapK} and \ref{mapJ} we show the maps of their values for 
three Liller~1 images acquired under different seeing conditions in the $K_s$ and $J$ bands, respectively.
The color coding is the same in both figures to allow a direct comparison.
As expected, better performance are obtained in better seeing conditions and closer to the guide star asterism,  
where AO corrections are more efficient, thus yielding smaller values of FWHM and higher values of the SR and EE. 
It is also interesting to note that the measured counts of the five laser guide stars, as reported in the header of the used images, 
can vary up to a factor of two, thus possibly contributing to some of the observed gradient.

\subsection{GeMS/GSAOI {\it versus} HST/ACS performance}
\label{hstsec}

In verifying the potentiality of ground-based MCAO-assisted imagers to obtain accurate photometry in dense stellar fields, it is 
very interesting to compare the performance of the GeMS/GSAOI system with those 
of HST. However, to perform a meaningful comparison, it is necessary to probe wavelength ranges where similar diffraction limits are
expected between the two telescopes. As the primary mirror of HST is about three times smaller ($\sim 2.4$m) 
than that of the Gemini South telescope, the diffraction limit expected for Gemini $J$ and $K_s$ images ($0.04\arcsec - 0.07\arcsec$), 
is obtained at optical wavelengths with HST. 

We used two short exposures ($t_{exp}=15$ sec) of NGC~6624 taken with 
Advanced Camera for Surveys/Wide Field Camera (ACS/WFC) onboard HST, in the F606W and F814W bands 
(Prop: 10775; PI: Sarajedini). 
A sample of about 200 high signal-to-noise and isolated stars 
have been selected to compute average FWHM, SR and EE and
their dispersions around the mean, by using the same analysis as for GeMS/GSAOI images.

We obtain average values of FWHM of 82 mas and 86 mas for the F606W and F814W, respectively. While the average FWHM for the F814W
is consistent with the nominal diffraction limit of HST at these wavelength ($\sim$ 85 mas), the FWHM in the F606W is significantly
larger (by $\sim30\%$) than the nominal diffraction limit ($\sim$ 63 mas), 
but this is somewhat expected, since at these wavelengths the limiting factor is 
the undersampling of the PSF. 
We find also that the overall variation of the FWHM along the entire ACS FOV ($\sim 200\arcsec\times 200\arcsec$) is $\sim8\%$.
This value is consistent with a $\pm 10\%$ variation estimated by using a significantly larger data-set by \citet{and06}.
For the same stars  we estimated $SR\sim50\%$ and $\sim65\%$ for the F606W and F814W, respectively, and
$\sigma_{SR}<10\%$ for both filters. Moreover, we find $EE\sim55\%$ and $\sigma_{EE}\sim10\%$ for both filters. These
latter values are consistent with those estimated by Sirianni et al. (2005; Table~3) within a comparable aperture of 
2$\times$FWHM (corresponding to an equivalent circular radius between 50 and 100 mas) for white dwarf spectro-photometric 
standards located at the center of the two ACS/WFC chips. 

In Figure~\ref{hst} we plot the dispersion around the average FWHM, SR and EE values as a function of the corresponding average values
for the GeMS/GSAOI $J$ and $K_s$ images, as well as for the ACS/HST F606W and F814W ones.
For sub-arcsec seeing conditions, GeMS/GSAOI 
delivers images with comparable or even better PSF FWHMs than ACS and also similar uniformity over the FOV
(at least in the $K_s$ band).
At variance, both the SR and the EE are in most cases lower 
than the corresponding values of ACS, 
while their variation over the FOV is comparable (around $10\%$).
Only in the best seeing conditions and in the $K_s$ band, GeMS/GSAOI can reach EE values comparable 
with those delivered by HST/ACS.

In Figure~\ref{6624map} 
we show the FWHM, SR and EE maps for three NGC~6624 
images in the $K_s$, $J$ and F814W filters, respectively.
The color coding is the same of Figures~\ref{mapK} and \ref{mapJ} for a direct comparison.
As in the case of Liller~1, GeMS/GSAOI 
delivers better performance 
(i.e. smaller FWHM and higher SR and EE) in the surrounding of the guide star asterism.
HST-ACS provides very uniform FWHMs over the entire FOV. SR and EE improve smoothly 
with increasing the radial distance from the center, 
as expected since crowding decreases.
The stars in the very central region of the cluster (white circular area in the bottom panels of Figure~\ref{6624map}) 
were not used to sample the PSF FWHM, SR and EE in 
the HST images, because measurements are quite uncertain due to the prohibitive crowding. 
However, it was possible to measure them  
in the GeMS/GSAOI NIR images (the sampled FOV is indicated as a black square in Figure~\ref{6624map}), 
since in that case crowding by resolved stars is less severe than in the HST images, 
due to a combination of a slightly higher spatial resolution and a 
lower sensitivity to faint stars, which only contribute in the form of unresolved stellar background.

\section{GeMS/GSAOI astrometric performance}

High-precision astrometry is crucial for many science cases in modern astrophysics.
In the study of globular cluster stellar populations, precise astrometry is required to measure proper motions and 
obtaining precious information on the contamination by field stars and on the internal kinematics (see for example 
\citealt{wat15}; \citealt{rich13}; \citealt{bel15}).
To measure stellar proper motions one needs to derive position displacements between two (or more) epochs.
However in virtually all available instruments, star displacements are not only due to ``real'' star motions, but
they are also the result of instrumental effects (distortions) that alter artificially 
the position of stars and that need to be modeled to obtain
highly accurate astrometric solutions.
In AO-assisted imagers, these distortions are not only of geometric nature, but they can
depend also on other factors (like for example anisoplanatism), although in the following we will always refer to them as 
``geometric distortions'' (GDs).
A successful approach to model GDs has been proposed for the first time by \citet{and03},
who found a distortion solution for the HST Wide Field Planetary Camera 2.
In recent years, other similar works aimed at measuring the GD of 
the ACS and the Wide Field Camera 3 \citep{bel09,bel11,bel14} onboard HST 
or of ground-based imagers such as LBT/LBC, ESO/WFI, VLT/HAWK-I, VISTA/VIRCAM 
\citep{and06,yad08,bellini09,bel10,lib14,lib15},
have been published.

An analysis of the internal astrometric performance of GeMS/GSAOI system has been recently presented by \citet{nei14b}.
They find that for single-epoch, well populated undithered images, an internal astrometric error of $\sim 0.2$ mas can
be achieved for well exposed images ($t_{\rm exp}>1$ min). On the contrary, for multi-epoch observations, an
additional systematic error of $\sim 0.4$ mas should be considered. According to the authors this is likely due
to time-variable distortion induced by gravity instrument flexure.

In this paper, we attempt to obtain the first formal analytic solution to the GDs of
GeMS/GSAOI for the $J$ and $K_s$ filters and an analysis of the absolute astrometric performance of the system. 
To this aim, we closely followed the approach described in \citet{and03} and
\citet{bel09}. 
We used the single epoch dithered (by $\sim3\arcsec$) images (14 in $K_s$ and 13 in $J$, see Table~\ref{tab1}) available for NGC~6624.
Indeed this is an ideal dataset for this goal for several reasons:
{\it i}) during the observing night, the atmospheric conditions were good and quite stable (average seeing $\approx$ 0.65\arcsec, 
airmass close to one);
{\it ii}) dithering allows to analyze both {\it dynamic} and {\it static} distortions of the camera \citep{nei14b};
{\it iii}) a distortion-free catalog of the cluster stars to be used as reference is available.

\subsection{A geometric distortion solution}

The most straightforward way to solve for the GD for a given instrument 
is to compare the instrumental positions of stars in that instrument with the corresponding ones in a 
distortion-free reference catalog, so that 
information about distortions can be directly derived from the stellar positional residuals.
For NGC~6624 we used as reference system the ACS catalog published by \citet{sar07}.
In this catalog the positions are corrected for GD effects 
using the solutions by \citet{and03} and \citet{meu03}. It covers a large enough FOV to entirely include the GeMS/GSAOI data-set 
and it samples the entire magnitude range probed by the GeMS/GSAOI images for NGC~6624. 

We combined the ACS catalog with data obtained with the WFC3 UVIS channel within the {\it HST UV Legacy survey of globular clusters} 
(Prop: 13297, PI: Piotto; see \citealt{pio15} for a description of the data-set) and derived relative proper motions by using 
the approach described in \citet{massari13}
and \citet{dalex13}. Only stars in common between the ACS and the WFC3 catalogs and with proper motions $(dx, dy)<0.1$ mas/yr
(corresponding to 0.002 pixel/yr) have been selected to build the {\it master catalog} and derive the GD solution.
Such a selection guarantees that the stars in the {\it master catalog} are cluster members and are virtually stationary within the proper
motion uncertainties, which correspond to $\Delta v\sim 3-4$ Km/s respect to the cluster systemic velocity at the distance of NGC~6624.
It is important to stress here that the adopted selection corresponds to the typical proper motion error 
for well measured stars obtained with similar data-sets.
Hence more restrictive criteria have no significant impact on the ``quality'' of the {\it master catalog}, but their main effect is 
on the sample size.   

We then matched it with the GeMS/GSAOI catalog 
of NGC~6624 described in \citet{sar16} and found $\sim 7500$ stars in common covering the magnitude range $13<K_s<19$.
The average crowding in this FOV is $\sim13$ stars/arcsec$^2$ at $K_s<20.3$ mag.
The ($V, V-I$) and ($K_s, J-K_s$) color-magnitude diagrams of the {\it master catalog} stars in
common with the GeMS/GSAOI catalog of NGC 6624 presented in \citet{sar16} are plotted in Figure~\ref{cmd}.
 
GDs have been computed for each chip, independently, taking as reference center in each chip the position
$(x_{0}, y_{0})_{k}$ = (1024, 1024) in raw pixel coordinates, where the index $k$ = 1, 2, 3, 4  
indicates the considered chip. In fact, obtaining a separate solution for each chip, 
rather than one that uses a common center of the distortion in the FOV,
allows a better handle of potential individual detector effects.

The main steps of the procedure can be summarized as follows \citep[see also][for more details]{bel09}.
\begin{itemize}
\item We conformally transformed\footnote{A conformal transformation, also called Helmert transformation, allows to switch 
from one reference system to another, through a change of scale, a rotation and two rigid shifts along x and y axes, respectively.} 
each {\it i} -- star coordinates in the {\it master catalog} $(X_i^{master}, Y_i^{master})$ into pixel coordinates of each 
dithered GeMS/GSAOI {\it j}--image.
We then cross-correlated these nominal positions with those actually measured in each GeMS/GSAOI {\it j}--image,
thus generating pairs of positional residuals: 

\begin{equation}
\Delta x_{i,j,k} = x_{i,j,k} - (X_i^{master})^{T_{j,k}}
\end{equation}

\begin{equation}
\Delta y_{i,j,k} = y_{i,j,k} - (Y_i^{master})^{T_{j,k}}
\end{equation}
\\
 
The set of positional residuals ($\Delta x_{i,j,k}, \Delta y_{i,j,k}$) 
of the stars in the raw images as a function of the $x_{i,j,k}$ and $y_{i,j,k}$ coordinates,  
determines the amount of GD and its spatial distribution in each chip.
\item To create a GD map, we divided each chip (2048$\times$2048 pixels) into grids of (16$\times$16) cells of (128$\times$128) 
pixels each, in order to have sufficient statistics but also spatial resolution.
In each cell we estimated the following parameters\footnote{The average values of these quantities have been 
obtained by applying a 3$\sigma$-rejection.}: $\overline{x}_{m,k}$, $\overline{y}_{m,k}$, $\overline{\Delta {x}}_{m,k}$, 
$\overline{\Delta {y}}_{m,k}$ and $P_{m,k}$, where {\it m} = 1, 256 is the cell reference number in our grid.
$\overline{x}_{m,k}$ and $\overline{y}_{m,k}$ are the average positions of all the stars in each grid cell, 
$\overline{\Delta {x}}_{m,k}$ and $\overline{\Delta {y}}_{m,k}$ the average positional residuals, 
while $P_{m,k}$ indicates the number of stars for each cell.
\item We represented our GD solution with a third-order polynomial (we omitted {\it i, j} and {\it k} indexes for simplicity):

\begin{equation}
\delta x = a_1\tilde{x}+a_2\tilde{y}+a_3\tilde{x}^2+a_4\tilde{x}\tilde{y}+a_5\tilde{y}^2+a_6\tilde{x}^3+a_7\tilde{x}^2\tilde{y}+a_8\tilde{x}\tilde{y}^2+a_9\tilde{y}^3
\end{equation}

\begin{equation}
\delta y = b_1\tilde{x}+b_2\tilde{y}+b_3\tilde{x}^2+b_4\tilde{x}\tilde{y}+b_5\tilde{y}^2+b_6\tilde{x}^3+b_7\tilde{x}^2\tilde{y}+b_8\tilde{x}\tilde{y}^2+b_9\tilde{y}^3
\end{equation}
\\

We verified that a larger number of degrees 
of freedom did not significantly improve our solution.
In this system, $\tilde{x}$ and $\tilde{y}$ indicate the positions of individual stars 
relative to the central pixel $(x_{0}, y_{0})_{k}$ = (1024, 1024) of each chip, 
and $a_{0,k}$..$a_{9,k}$ and $b_{0,k}$..$b_{9,k}$ are the 18 coefficients that we need to determine.

To do this, we performed a linear least-square fit of the 256 data points 
in each grid cell, which actually means solving a matrix system.
\item After the first GD solution, we determined the distortion corrected positions 
($x^{corr}_{i,j,k}$, $y^{corr}_{i,j,k}$) as the observed positions 
($x_{i,j,k}$, $y_{i,j,k}$) plus the distortion corrections ($\delta x_{i,j,k}$, $\delta y_{i,j,k}$):
 
 \begin{equation}
y^{corr}_{i,j,k} = x_{i,j,k} + \delta x(\tilde{x}_{i,j,k},\tilde{y}_{i,j,k})
\end{equation}

\begin{equation}
y^{corr}_{i,j,k} = y_{i,j,k} + \delta y(\tilde{x}_{i,j,k},\tilde{y}_{i,j,k})
\end{equation}

\item The procedure was iterated more than 30 times for chip by applying at each iteration, only half of the correction in order to 
avoid convergence problems, until the difference in the positional residuals 
($\Delta x_{i,j,k}$, $\Delta y_{i,j,k}$ vs $x_{i,j,k}$, $y_{i,j,k}$) from one iteration to the following one became 
negligible (in other words, when the $\chi_{iterN}^2$ $\approx$ $\chi_{iterN+1}^2$).
\end{itemize}

The coefficients $a_{q,k}$ and $b_{q,k}$ ($q=1...9$) of the final GD solution for the four chips 
are given in Table~\ref{coeff_K} for the $K_s$ filter, and in Table~\ref{coeff_J} for the $J$ one.

Figures~\ref{gdmk} and \ref{gdmj} show the GD map and the residual trends 
of uncorrected star positions for the four chips of the GSAOI camera, in the $K_s$ 
and $J$ filters, respectively. 
In both cases, the size of the residual vectors is magnified by a factor of 10. 
Residual vectors connect the uncorrected average positions within each grid cell to the corrected ones.
We also show the overall trend of the positional residuals ($\Delta x_{i,j,k}$ vs $x_{i,j,k}$; $y_{i,j,k}$) 
and ($\Delta y_{i,j,k}$ vs $x_{i,j,k}$; $y_{i,j,k}$). 
These trends are quite similar/symmetric in the four chips, with a maximum amplitude of $\sim 30$ pixels both along the x and y axis.\\
In Figures~\ref{gdrk} and \ref{gdrj} we show the final residuals of the star positions in $K_s$ and $J$ filters, respectively, 
after applying our GD solution (residuals vectors are magnified by a factor 5000). 
Our GD solution allowed to linearize the residual trends in each chip and to reach 
an astrometric accuracy of about 0.07 pixels corresponding to $\sim 1.5$ mas.
In this respect it is worth recalling that, given the proper motion selection ($0.1$ mas/yr) adopted to build the {\it master catalog}
and that the GeMS/GSAOI images have been acquired about 7 yr after the first HST epoch, the total contribution to the final GD error budget due to
proper motions can be as large as $\sim0.7$ mas in this case.

As shown by \citet{nei14b}, the astrometric accuracy of ground-based instruments 
in general, and assisted by AO-systems in particular (e.g. GeMS/GSAOI), depends on 
a few major factors, namely:
{\it i}) the PSF shape and variability in the FOV;
{\it ii}) the atmospheric conditions (i.e. the seeing at the observing airmass);
{\it iii}) the brightness of the three natural guide stars and their asterism;
{\it iv}) the crowding of the observed field;
{\it v}) the exposure time.

Because of these important factors, our GD solution based on the observations of NGC~6624 should be 
further tested on other stellar fields with different crowding and 
acquired with different atmospheric conditions, asterisms and exposure times and with 
available high resolution astrometric reference catalogs (likely from HST imaging). 
While the general formal solution would likely be still valid at a first order approximation, 
data-sets obtained under different observing conditions may yield different coefficients.
Unfortunately, our observations of Liller~1 cannot be used for such a test since for this very reddened cluster 
an astrometric reference catalog based on high resolution data is not available. 
 
However, we have verified that the GD residuals (after correction) obtained for NGC~6624, 
are quantitatively compatible with those obtained for NGC~6681 \citep{massari16} observed under 
different seeing conditions and NGS asterism.
Our results are also qualitatively consistent with those obtained by Ammons et al. (2016) for the globular cluster NGC~1851, 
although a more
quantitative comparison is not possible because the authors do not provide enough details.
 
It is also important to stress that the analysis described in this Section has a general relevance, since it provides 
the mathematical formalism to correct for GDs and it can be thus effectively applied to any photometric data-set.

\section{Conclusions}

The PSF tests performed on the GeMS/GSAOI $J$ and $K_s$ images of the central region of two high-density 
globular clusters have provided very interesting results.

The ground-based MCAO-assisted imager GeMS/GSAOI instrument at Gemini South observatory provides a unique and powerful 
facility to derive accurate stellar photometry at nearly the diffraction limit spatial resolution, at least in the $K_s$ band where 
AO correction is more efficient.
Uniform (at a level of $\sim$10\%) PSF over $1\arcmin -2\arcmin$ FOV with up to $50-60\%$ EE within 2 FWHM can be
obtained in good seeing conditions. These performance are comparable with those delivered by HST imagers at optical
wavelengths.   

Following the same strategy adopted for other imagers onboard HST and for wide field ground-based cameras, we 
were also able to compute GDs for GeMS/GSAOI and provide corrected images with an astrometric accuracy of $\sim$ 1.5 mas 
in a stellar field with a crowding of $\sim$13 stars/arcsec$^2$ at $K_s\le$ 20.3 mag,  
thus demonstrating that a ground-based MCAO-assisted imager at an 8m-class telescope 
can provide accurate NIR photometry and absolute astrometry for proper motion studies in very dense stellar fields, 
in some cases outperforming HST's capability, due to the higher spatial resolution.
HST with its ACS and WFC3 imagers remains somewhat unique in providing photometry of the faintest stars in less crowded regions.

Looking at the future facilities, complementarity can be foreseen also with JWST, that will provide the deepest photometry 
in the NIR over a few arcmin FOV at about the same spatial resolution of HST,
and ground-based MCAO-assisted imagers at 20-40m class telescopes, which will provide significantly higher spatial resolution 
at the same wavelengths of JWST but over a smaller FOV. 


\acknowledgements
We thank the anonymous referee for his/her useful comments and suggestions.
E.D. and S.S. thank Andrea Bellini and Benoit Neichel for useful discussions and suggestions.
L.O. acknowledges the PRIN-INAF 2014 CRA 1.05.01.94.11: 
``Probing the internal dynamics of globular clusters. The first
comprehensive radial mapping of individual star kinematics with the 
new generation of multi-object spectrographs'' (PI: L.
Origlia).
D.G., S.V. and R.E.C. gratefully acknowledge support from the Chilean BASAL 
Centro de Excelencia en Astrof\'isica y Tecnolog\'ias Afines (CATA) grant PFB-06/2007.
R.E.C. also acknowledges funding from Gemini-CONICYT for Project 32140007.



\newpage

\begin{table*}
\caption{Dataset properties.}
\label{tab1}     
\begin{center}
\begin{tabular}{llcccc}  
\hline\hline
Cluster & Date & \# J-exp & \# $K_s$-exp &$<$s(500nm)$>$ & $<$airmass$>$\\
\hline
Liller 1 & 20 April 2013 &  3 & 10 & 1.07&1.03\\ 
Liller 1 & 22 May 2013   &  0 &  5 &1.05 &1.44\\ 
Liller 1 & 24 May 2013   &  9 &  0 & 0.75 &1.018\\ 
NGC 6624 & 24 May 2013   & 13 & 14 & 0.66&1.02 \\ 
\hline
\end{tabular}
\end{center}
{\bf Note}: Average seeing values at zenith s(500nm) are in arcsec.
\end{table*}
\begin{table*}
\begin{center}
\caption{Selected tip-tilt guide stars.}
\label{tab2}     
\begin{tabular}{llllllll} 
\hline\hline
Star & $K_s$ & H & $J$ & $I$ & $V$ & $R$ & $fmag$ \\
\hline
{\it Liller~1} &&&\\
NGS~1 (\footnotesize{2MASS J17332197-3323043}) &  12.799  &  12.715  &  12.958  & --&--&--& 15.338\\
NGS~2 (\footnotesize{2MASS J17332484-3322502}) &  11.127  &  11.185  &  11.418  & --&--&--& 13.947\\
NGS~3 (\footnotesize{2MASS J17332609-3323129}) &  11.297  &  11.074  &  11.735  & --&--&--& 13.577\\
\hline
{\it NGC~6624} &&&\\
NGS~1 (\footnotesize{2MASS J18233752-3022018}) &  ~8.472  &  8.723&   9.730  & 11.3850 & 13.4520 & 13.094 & 13.793\\
NGS~2 (\footnotesize{2MASS J18234108-3022221}) &  ~8.987  &  9.202&  10.100  & 11.6230 & 12.9890 & 13.219 & 13.662\\
NGS~3 (\footnotesize{2MASS J18234052-3021392}) &  ~8.827  &  9.274&   9.899  & 12.8800 & 14.1040 & 13.776 & 10.889\\
\hline
\end{tabular}
\end{center}

{\bf Note}: Identification name and $K_s, H, J$ magnitudes from 2MASS; $V,I$ from \citet{sar07}; $R$ estimated from isochrones and $fmag$ (in the 579-642 nm spectral range) from the UCAC3 catalog.
\end{table*}

\begin{table*}[!htbp]
\begin{center}
\caption{Coefficients of the third-order polynomial for each chip, representing 
the final GD solution for the $K_s$ filter.\label{coeff_K}}
\footnotesize
\begin{tabular}{c c c c c c c c c c}
\\
\hline
\hline
Term(q) & Polyn. & a$_{q,[1]}$ & b$_{q,[1]}$ & a$_{q,[2]}$ & b$_{q,[2]}$ & a$_{q,[3]}$ & b$_{q,[3]}$ & a$_{q,[4]}$ & b$_{q,[4]}$\\
\hline
\\
1 & $\tilde{x}$                 & 7.2959  & -8.1224 & -8.2342 & -8.0700 & -8.6718 & 5.5598 & 7.7150 & 5.3724\\
2 & $\tilde{y}$                 & -8.6592 & -6.9948 & -9.5140 & 7.0969 & 5.8746 & 7.4350 & 6.7219 & -5.4231\\
3 & $\tilde{x}^2$             & 6.7348  & 0.0217 & 7.0562 & 0.0903 & 6.9963 & 0.0301 & 6.8232 & 0.1139\\
4 & $\tilde{x}\tilde{y}$     & 0.1646  & -0.0045 & 0.1435 & -0.0719 & 0.1301 & -0.0197 & 0.2890 & 0.1021\\
5 & $\tilde{y}^2$             & 6.6305  & 0.1600 & 6.7711 & 0.2774 & 6.7787 & 0.2803 & 6.7095 & 0.4254\\
6 & $\tilde{x}^3$             & 0.0688  & 0.0089 & -0.0635 & 0.0066 & -0.0855 & 0.0042 & 0.1567 & 0.0100\\
7 & $\tilde{x}^2\tilde{y}$ & 0.0251  & 0.0770 & 0.0941 & 0.0112 & -0.0543 & -0.0347 & 0.0113 & -0.0045\\
8 & $\tilde{x}\tilde{y}^2$ & -0.0922  & -0.0215 & 0.0010 & 0.0774 & -0.0600 & 0.0543 & -0.0449 & 0.0378\\
9 & $\tilde{y}^3$             & 0.0300  & 0.0824 & -0.0325 & -0.0003 & 0.0305 & 0.0243 & -0.0151 & 0.0544\\
\\
\hline
\end{tabular}
\end{center}
\end{table*}

\begin{table*}[!htbp]
\begin{center}
\caption{Coefficients of the third-order polynomial for each chip, representing 
the final GD solution for the $J$ filter.\label{coeff_J}}
\footnotesize
\begin{tabular}{c c c c c c c c c c}
\\
\hline
\hline
Term(q) & Polyn. & a$_{q,[1]}$ & b$_{q,[1]}$ & a$_{q,[2]}$ & b$_{q,[2]}$ & a$_{q,[3]}$ & b$_{q,[3]}$ & a$_{q,[4]}$ & b$_{q,[4]}$\\
\hline
\\
1 & $\tilde{x}$                 & 7.3378 & -8.0391 & -8.2090 & -8.1578 & -8.5896 & 5.6014 & 7.3445 & 5.3859\\
2 & $\tilde{y}$                 & -8.5374 & -6.8857 & -9.5506 & 7.1074 & 5.9348 & 7.4968 & 6.6005 & -5.4962\\
3 & $\tilde{x}^2$             & 6.6703 & 0.0347 & 7.0411 & 0.0730 & 6.9823 & 0.0685 & 6.7324 & -0.0230\\
4 & $\tilde{x}\tilde{y}$     & 0.1721 & 0.0946 & 0.1203 & -0.1116 & 0.1075 & -0.0342 & 0.3678 & 0.1387\\
5 & $\tilde{y}^2$             & 6.6176 & 0.1077 & 6.7432 & 0.2616 & 6.7769 & 0.2681 & 6.6951 & 0.2807\\
6 & $\tilde{x}^3$             & 0.0220 & -0.0157 & -0.0464 & 0.0125 & -0.1522 & -0.0251 & 0.2020 & -0.0321\\
7 & $\tilde{x}^2\tilde{y}$ & -0.0288 & 0.0233 & 0.1379 & 0.0417 & -0.1129 & -0.0846 & 0.0102 & -0.0558\\
8 & $\tilde{x}\tilde{y}^2$ & -0.2149 & -0.2188 & 0.0099 & 0.1537 & -0.1227 & 0.0069 & -0.0205 & 0.0358\\
9 & $\tilde{y}^3$             & 0.0112 & 0.0140 & 0.0088 & -0.0184 & 0.0092 & -0.0102 & -0.0273 & 0.0575\\
\\
\hline
\end{tabular}
\end{center}
\end{table*}

\begin{figure*}
\centering
\includegraphics[width=\hsize]{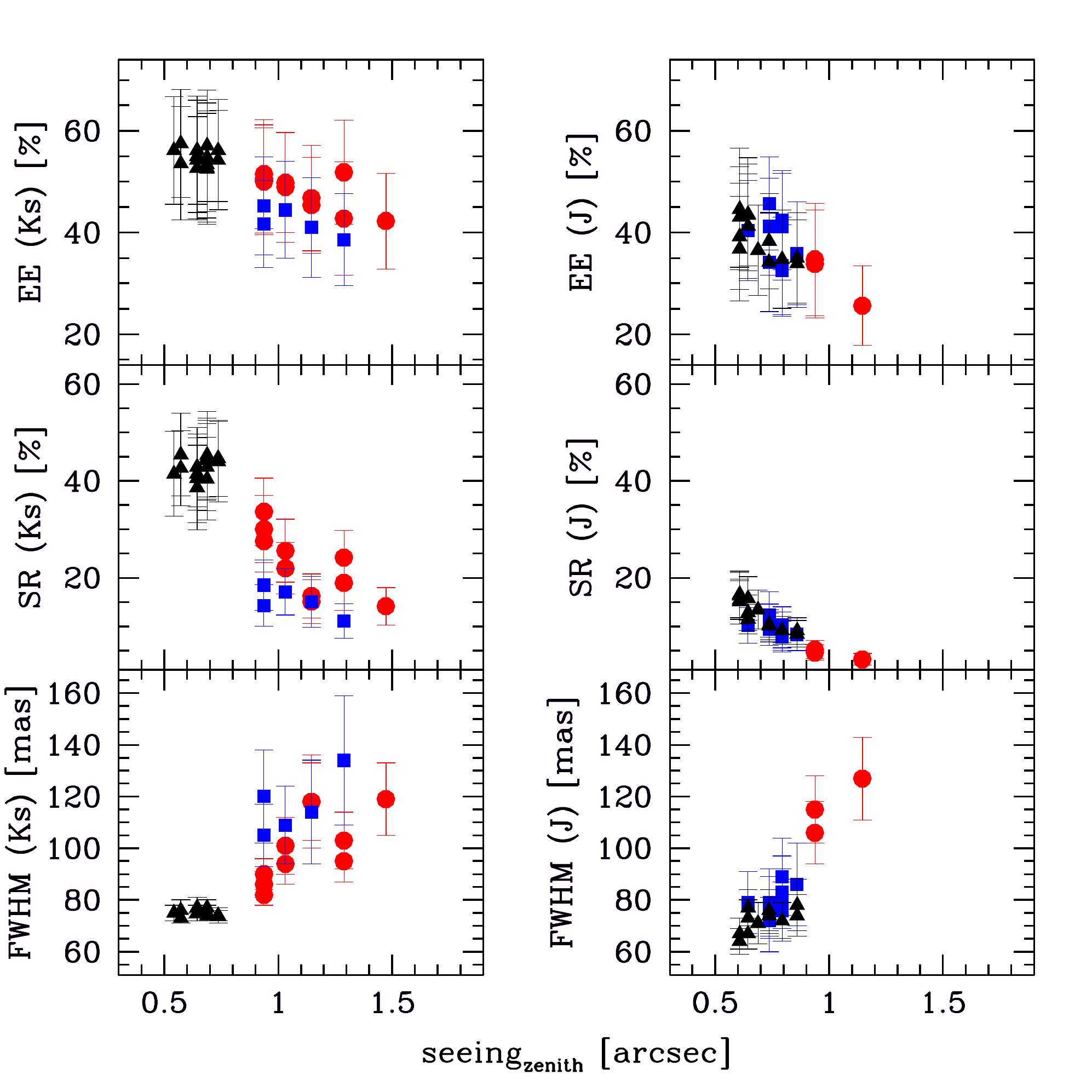}
\caption{Average FWHM, SR and EE values with varying the seeing at 500nm at the zenith.  
Left panels: measurements in the $K_s$ band, right panels: measurements in the $J$ band. 
Triangles refer to measurements of stars in NGC~6624, circles and squares refer to measurements of stars in Liller~1 
observed in two different nights, respectively (see Table~\ref{tab1}).}
\label{paramz}
\end{figure*}

\begin{figure*}
\centering
\includegraphics[width=\hsize]{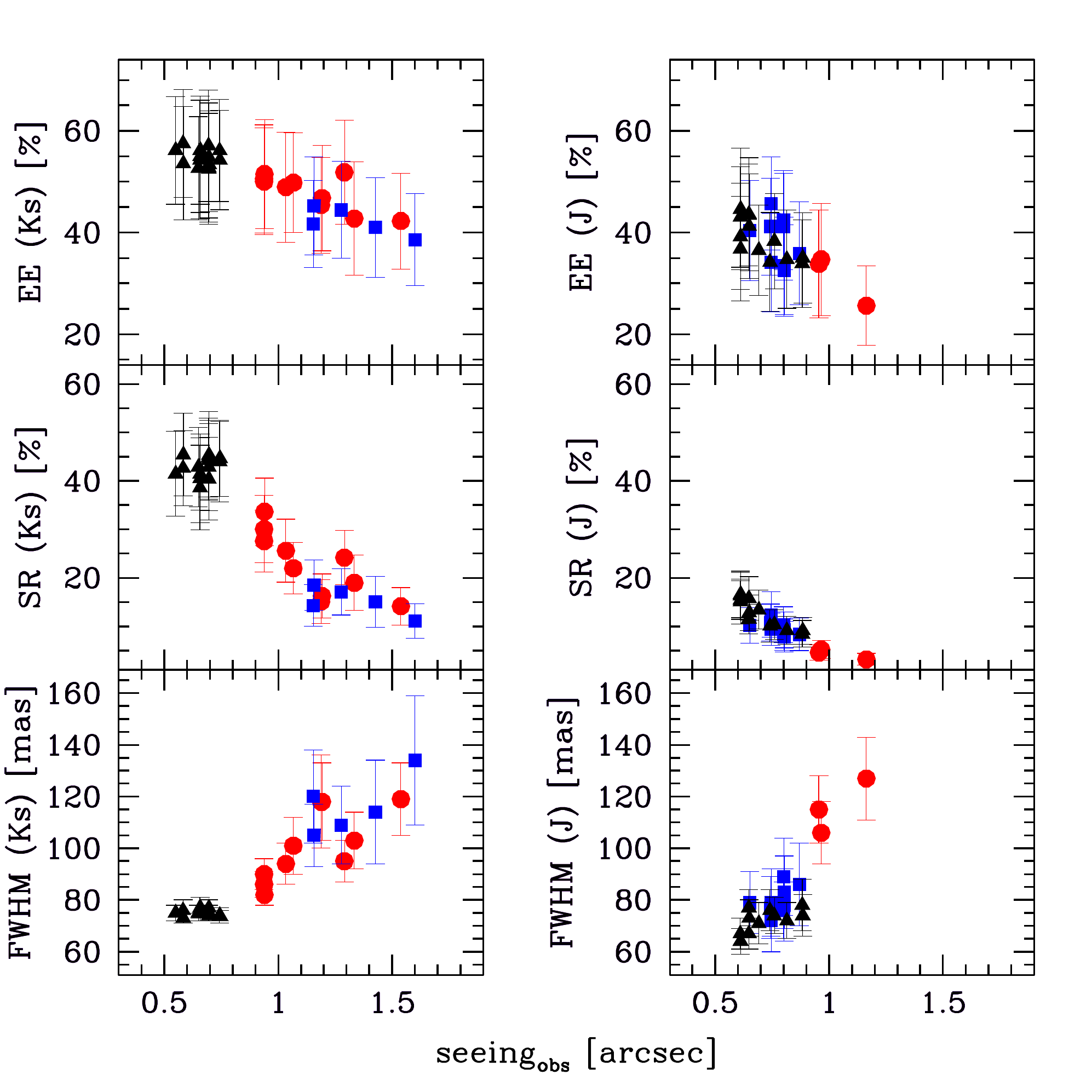}
\caption{Same as in Figure~\ref{paramz}, but for seeing at 500nm at the observing airmass.}
\label{parama}
\end{figure*}

\begin{figure*}
\centering
\includegraphics[width=\hsize]{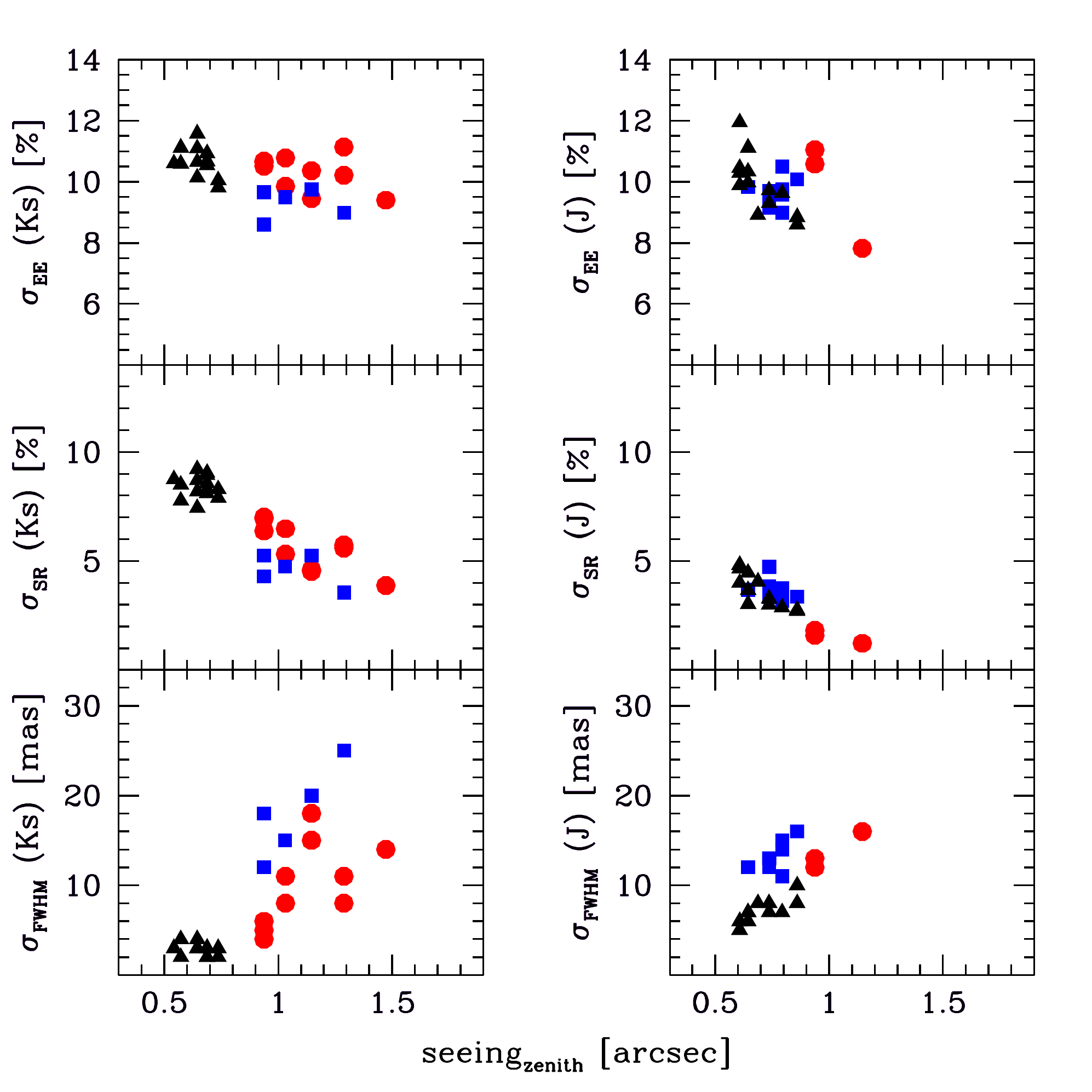}
\caption{Dispersion around the average FWHM, SR and EE values with varying the seeing at 500nm at the zenith.
Left panels: measurements in the $K_s$ band, right panels: measurements in the $J$ band. 
Triangles refer to measurements of stars in NGC~6624, circles and squares refer to measurements of stars in Liller~1 
as observed in two different nights, respectively (see Table~\ref{tab1}).}
\label{errz}
\end{figure*}

\begin{figure*}
\centering
\includegraphics[width=\hsize]{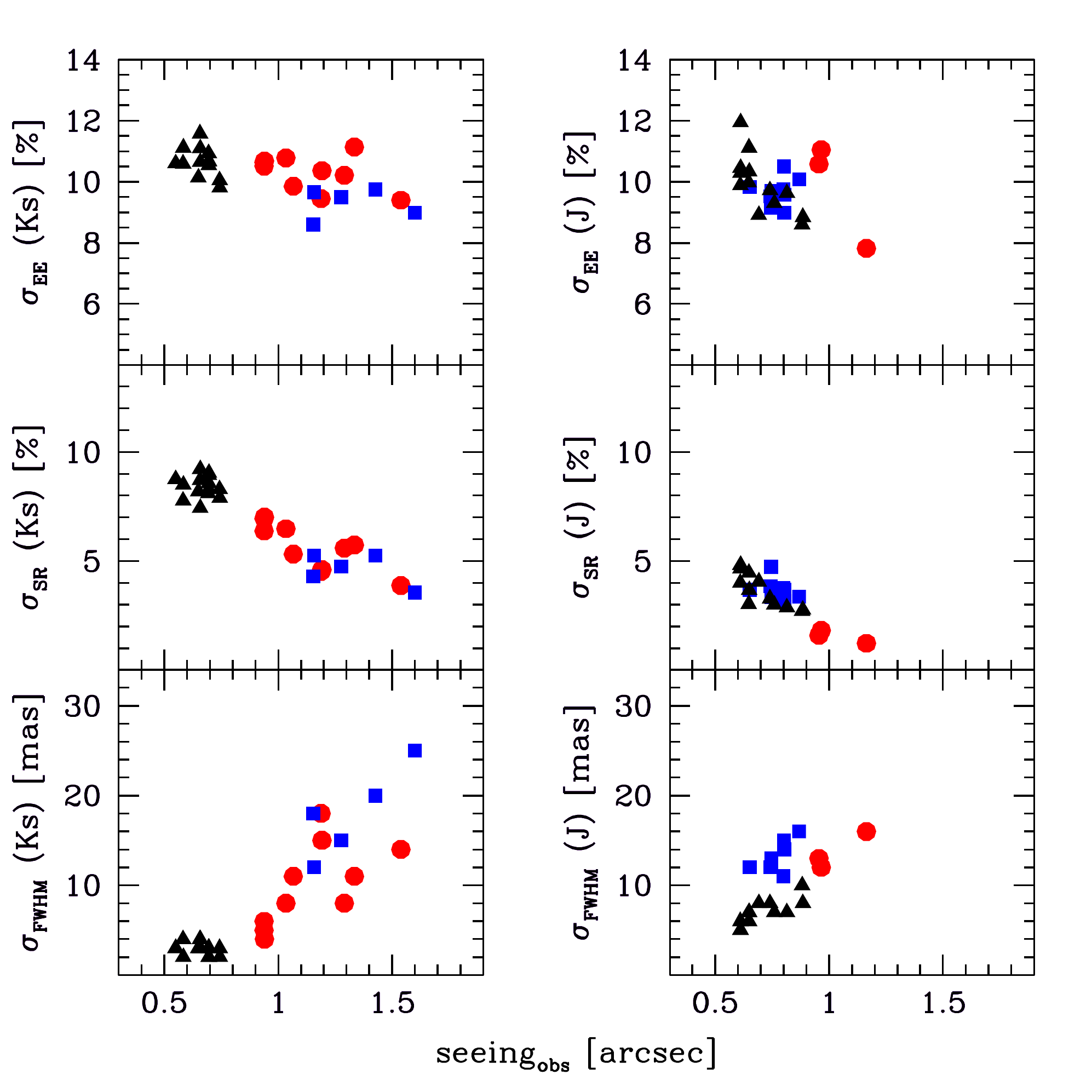}
\caption{Same as in Figure~\ref{errz}, but for seeing at 500nm at the observing airmass.}
\label{erra}
\end{figure*}
\begin{figure*}
\centering
\includegraphics[width=\hsize]{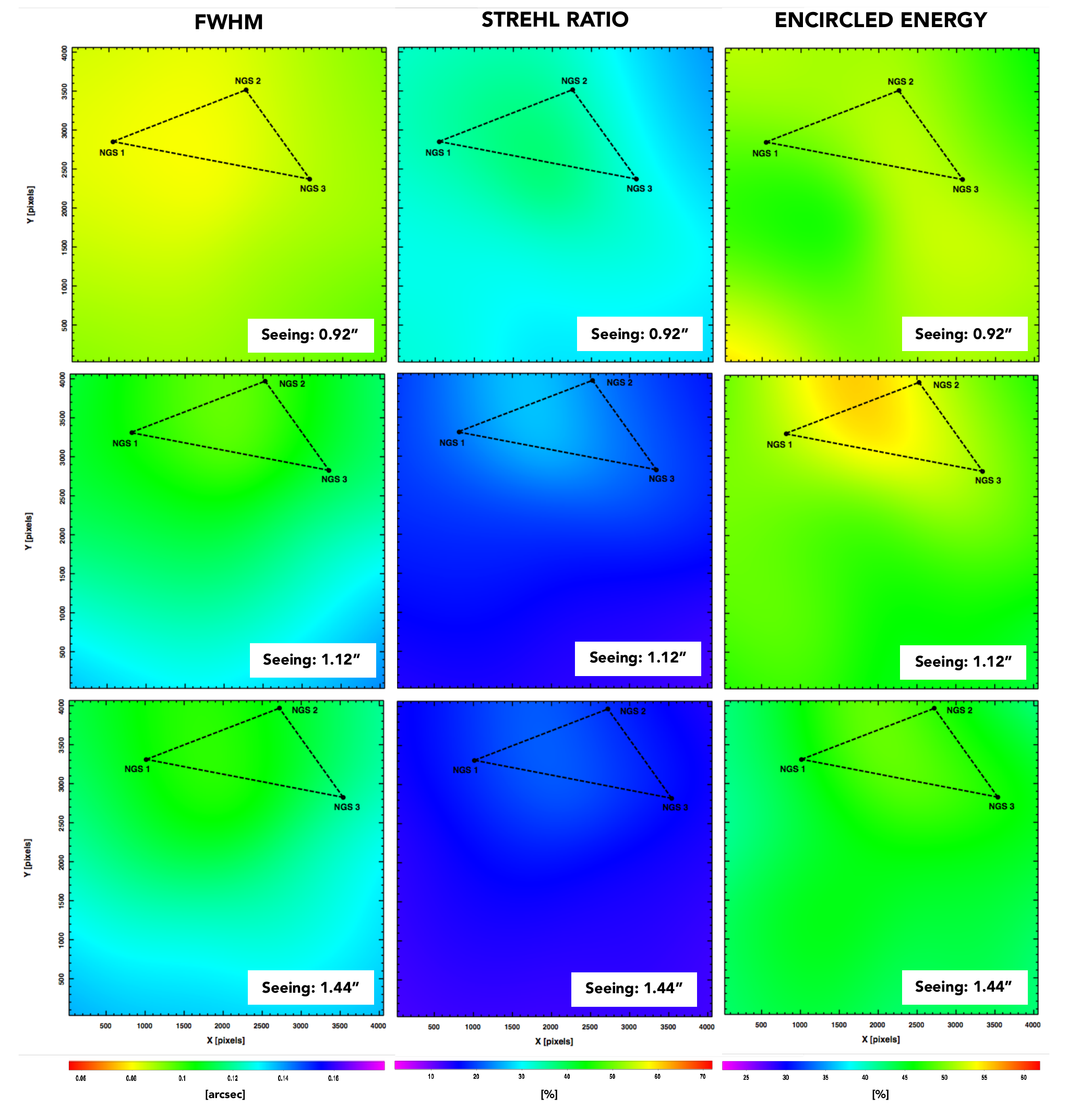}
\caption{FWHM (left panels), SR (middle panels) and EE (right panels) maps for
 three $K_s$ band images of Liller~1 acquired under different seeing conditions.
The triangle indicates the guide star asterism.
The quoted seeing values are at the zenith and at 500nm.
Color coding from magenta (worst) to red (best) is a performance indicator.}
\label{mapK}   
\end{figure*}
\begin{figure*}
\centering
\includegraphics[width=\hsize]{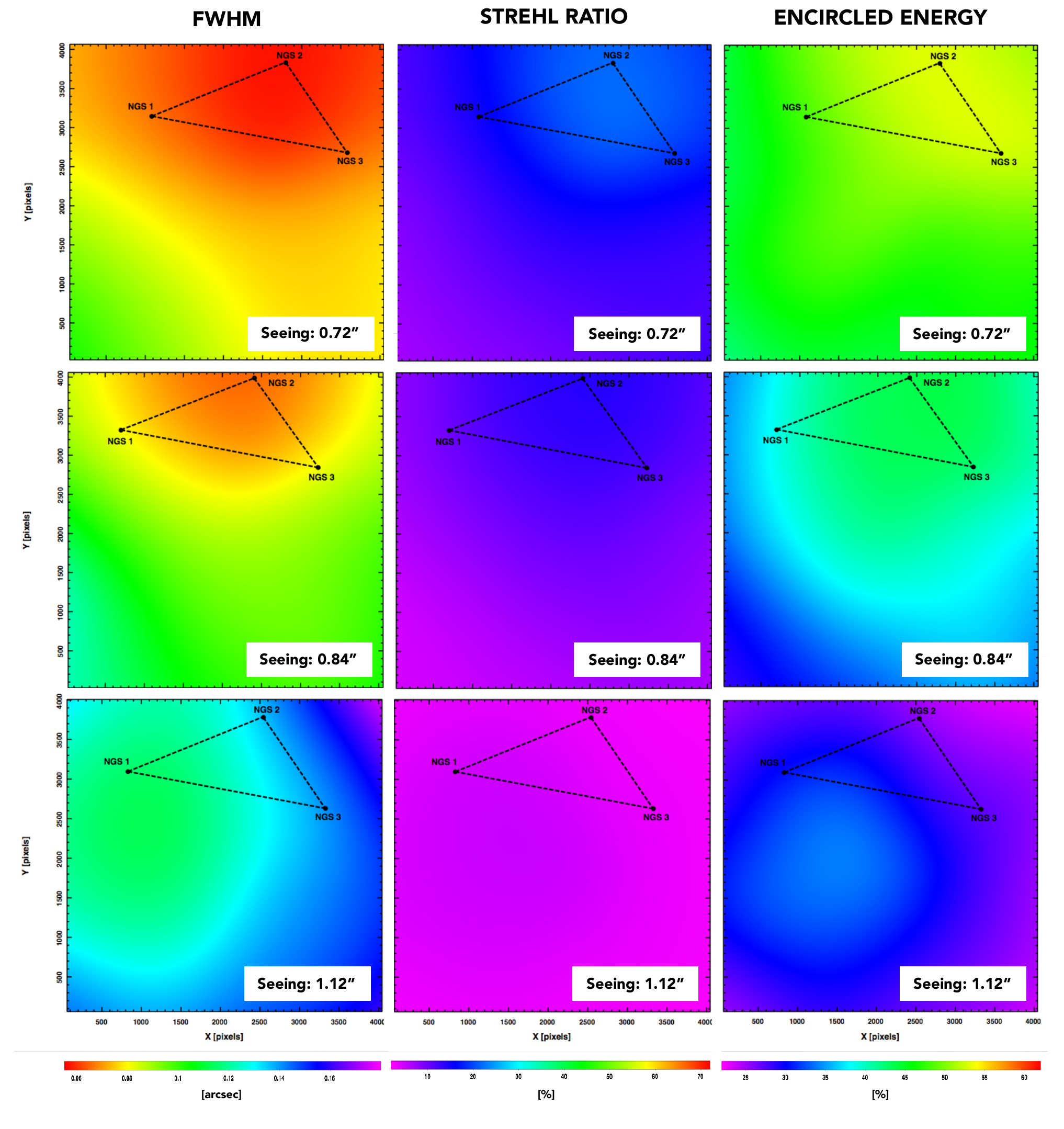}
\caption{FWHM (left panels), SR (middle panels) and EE (right panels) maps for
 three $J$ band images of Liller~1 acquired under different seeing conditions.
The triangle indicates the guide star asterism.
The quoted seeing values are at the zenith and at 500nm.
Color coding from magenta (worst) to red (best) is a performance indicator.}
    \label{mapJ}   
    \end{figure*}
  \begin{figure*}
  \centering
  \includegraphics[width=\hsize]{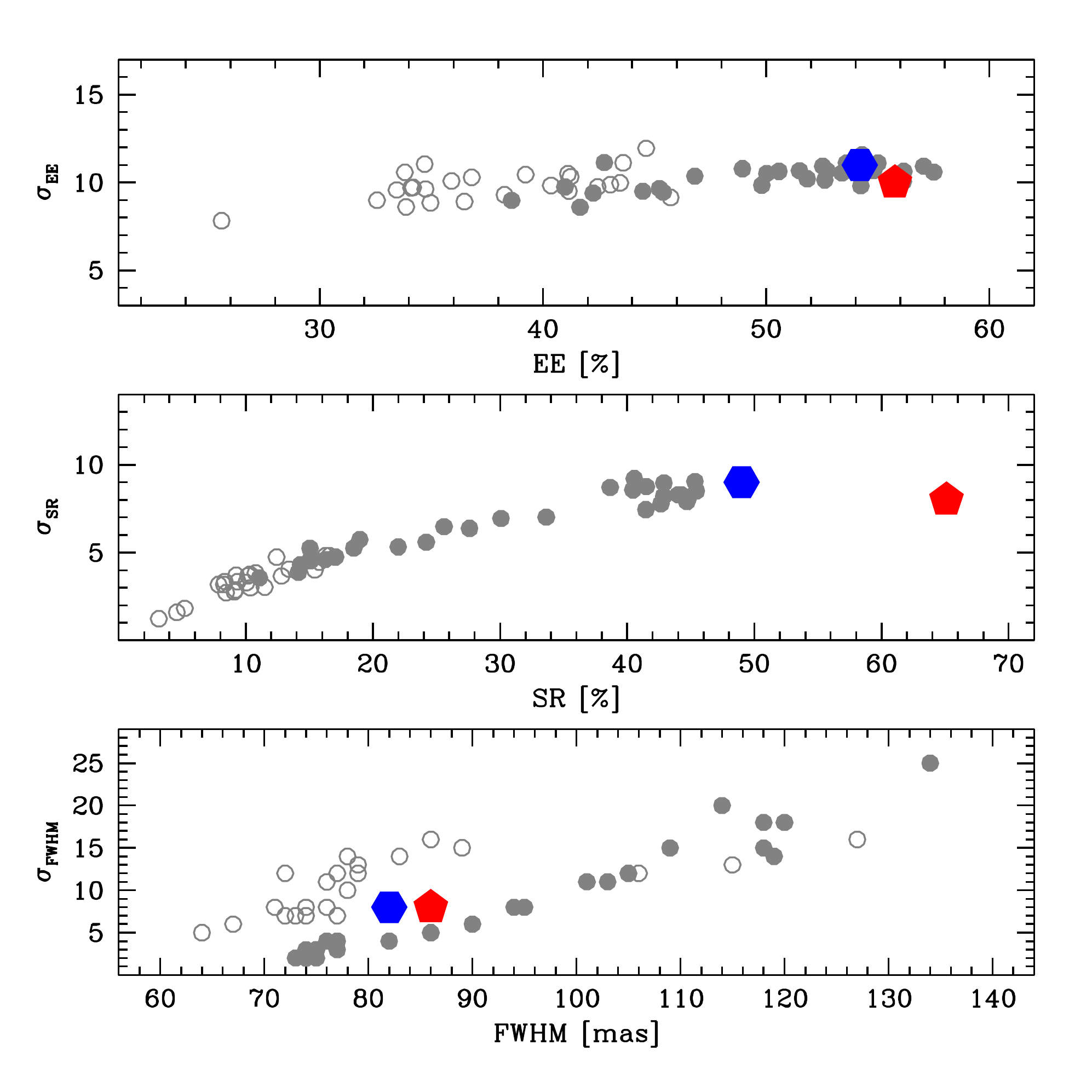}
   \caption{Dispersion around the average FWHM, SR and EE values as a function of the corresponding average values 
for the GeMS/GSAOI $J$ (open circles) and $K_s$ (filled circles) images, and the ACS/HST F606W (hexagon) and F814W (pentagon) 
ones.}
    \label{hst}   
    \end{figure*}
  \begin{figure*}
  \centering
  \includegraphics[width=\hsize]{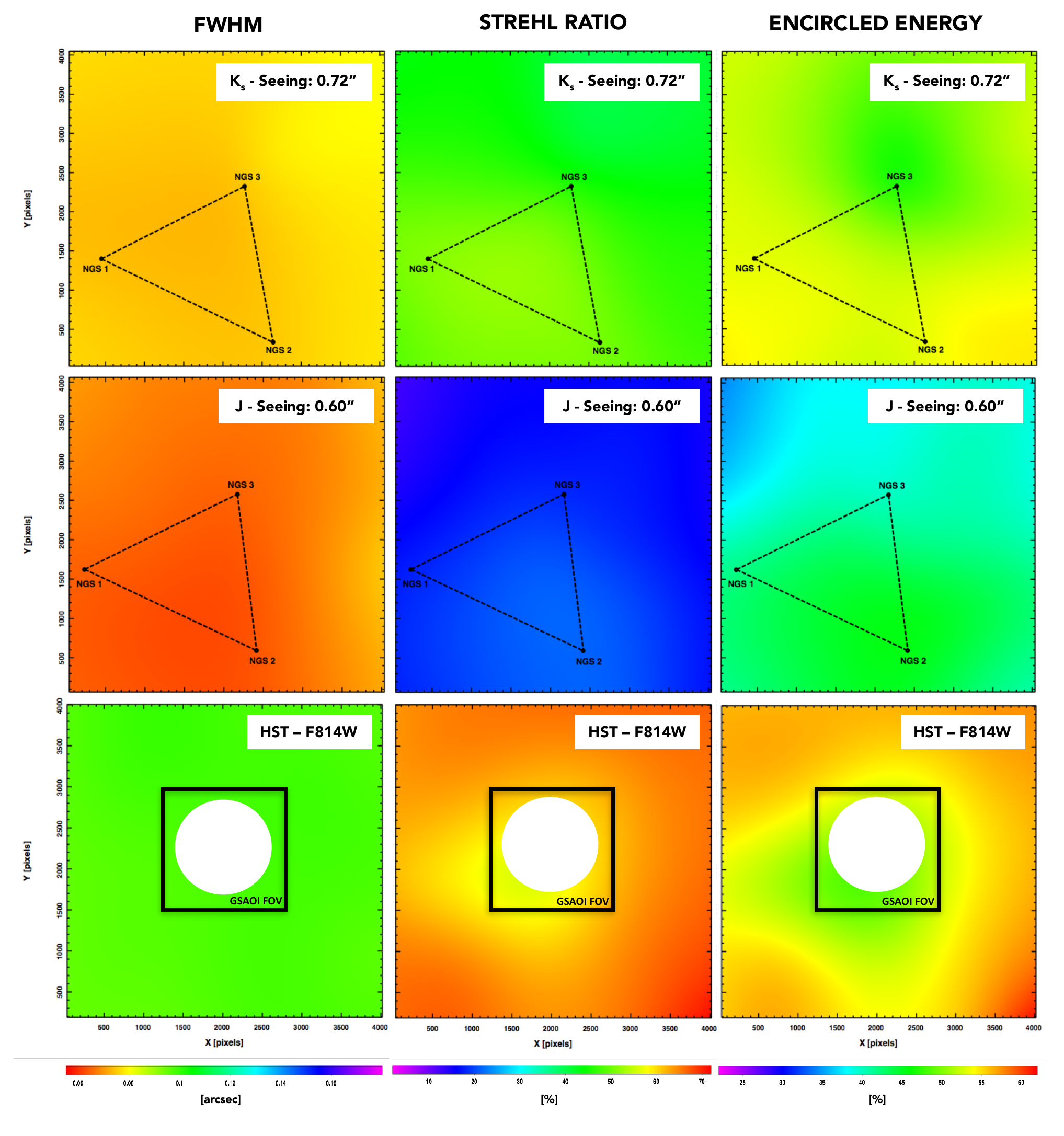}
   \caption{FWHM (left panels), SR (middle panels) and EE (right panels) maps for three NGC~6624 
images in the $K_s$ (top panels), $J$ (middle panels) and F814W (bottom panels) filters, respectively.
The triangle indicates the guide star asterism for the GSAOI ground-based images. The central, white area 
in the HST maps was excluded due to the prohibitive crowding, while the black square indicates the GeMS/GSAOI FOV.}
    \label{6624map}   
    \end{figure*}
  \begin{figure*}
  \centering
  \includegraphics[width=\hsize]{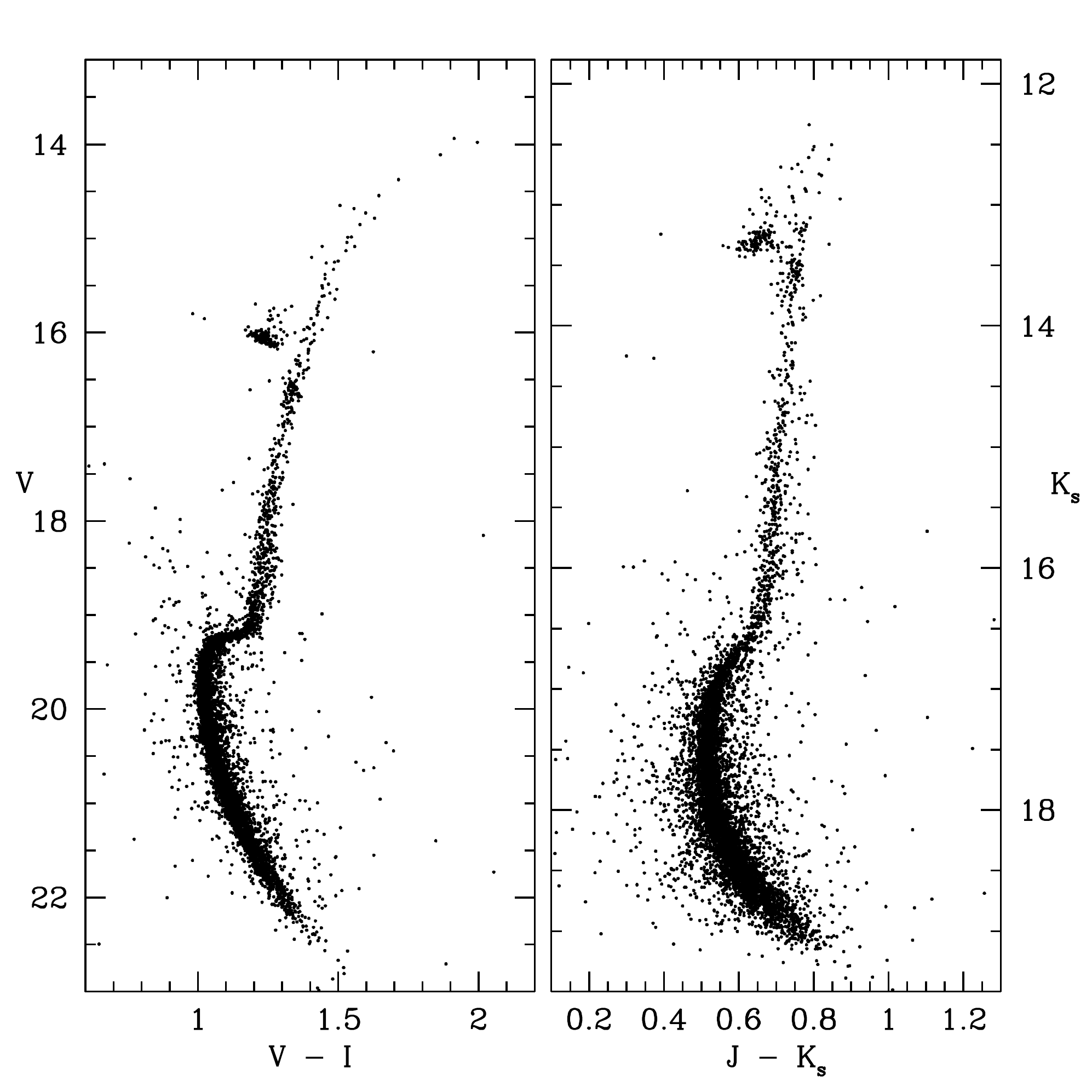}
   \caption{$V, V-I$ (left panel) and $K_s, J-K_s$ (right panel) CMDs of the stars in common between 
   the {\it master catalog} and the GeMS/GSAOI catalog of NGC 6624 by \citet{sar16}.} 
    \label{cmd}   
    \end{figure*}
\begin{figure*}[!htbp]
\centering
{\includegraphics[width=17cm, angle=0]{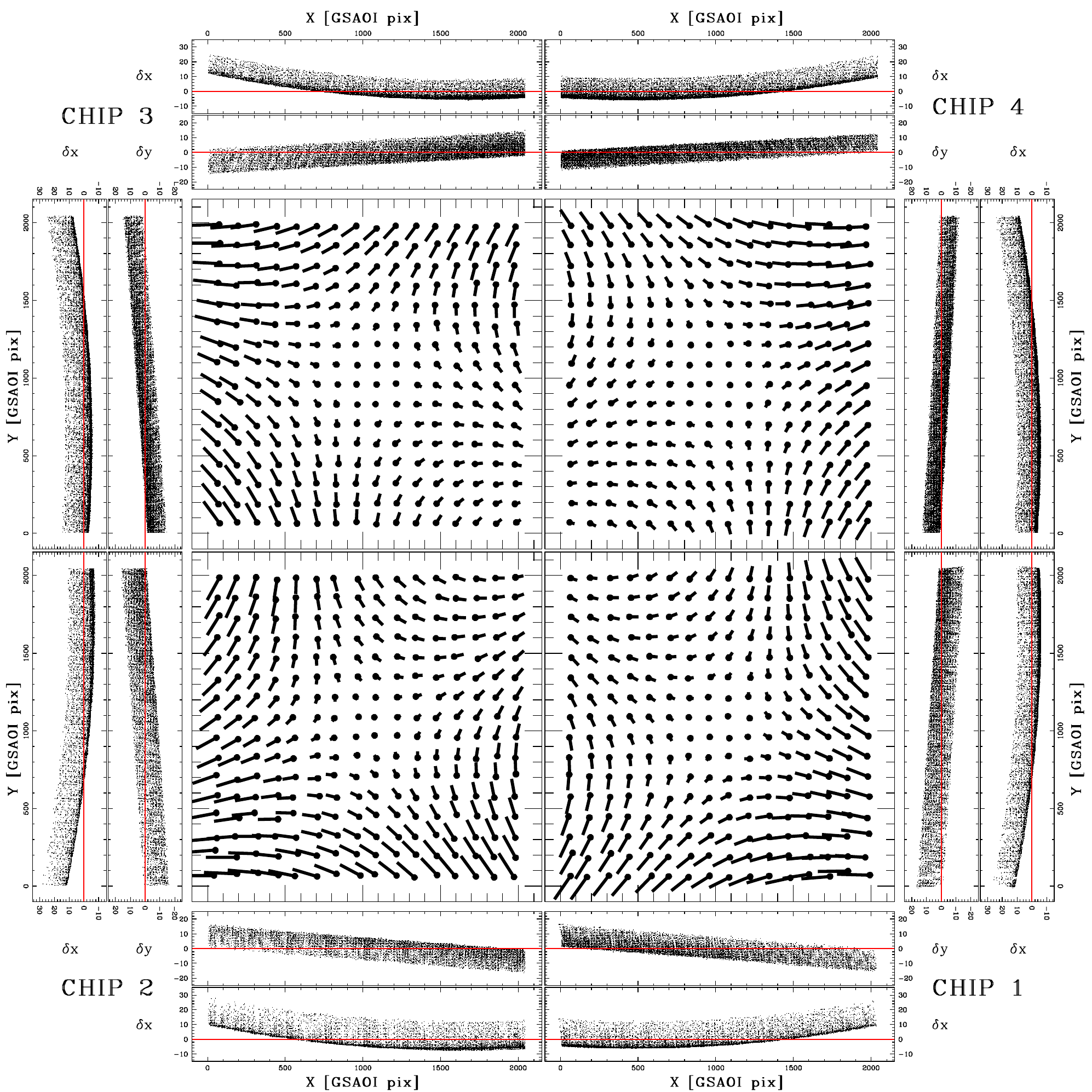}}
\caption{Geometric distortion map of the four chips of GSAOI camera, in the $K_s$ filter. 
Residual vectors are magnified by a factor of 10. 
For each chip, we also show individual residuals as function of {\it x} and {\it y} axes. Units are in GSAOI pixels.}
\label{gdmk}
\end{figure*}
\begin{figure*}[!htbp]
\centering
{\includegraphics[width=17cm, angle=0]{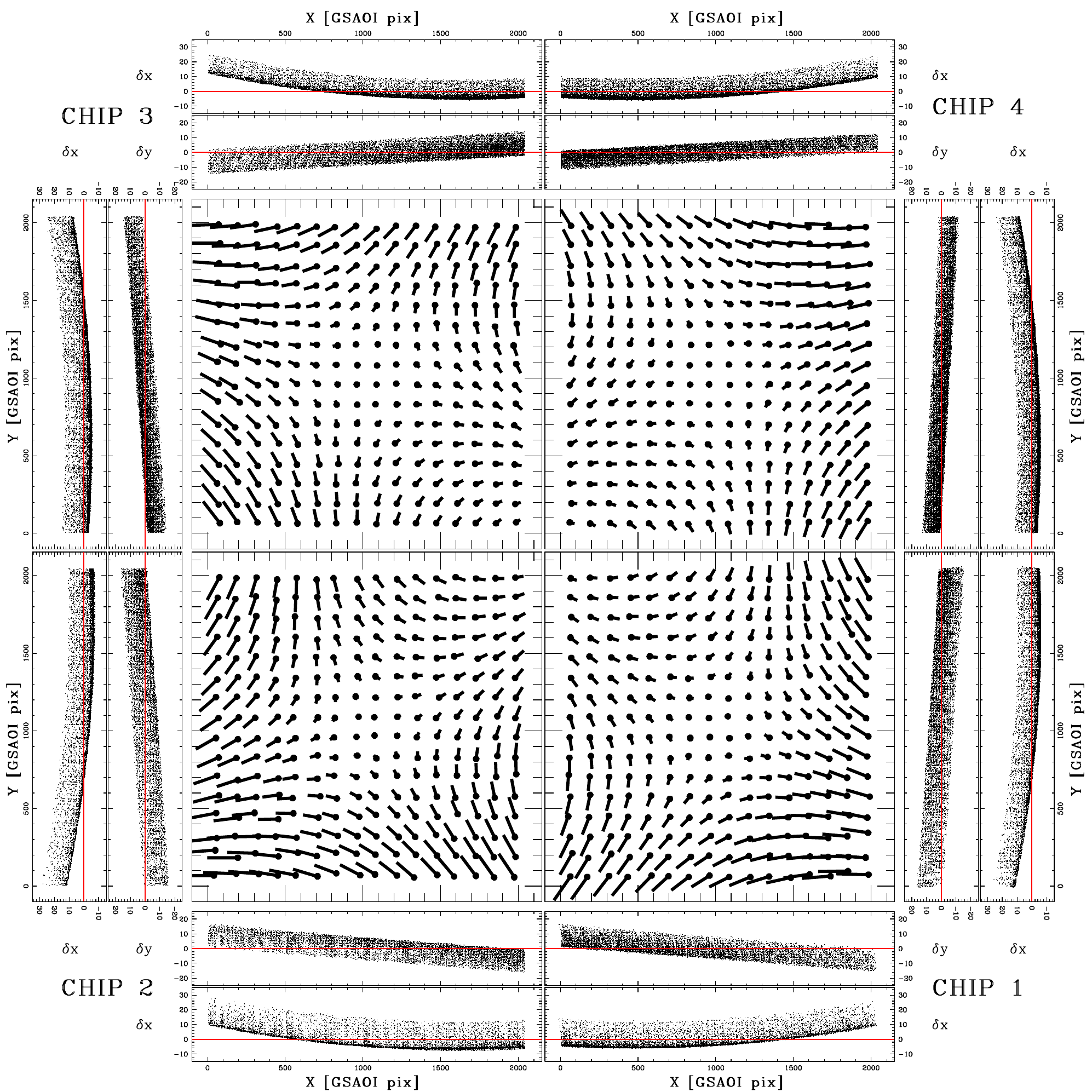}}
\caption{As in Figure~\ref{gdmk}, but for the $J$ band.}

\label{gdmj}
\end{figure*}
\begin{figure*}[!htbp]
\centering
{\includegraphics[width=17cm, angle=0]{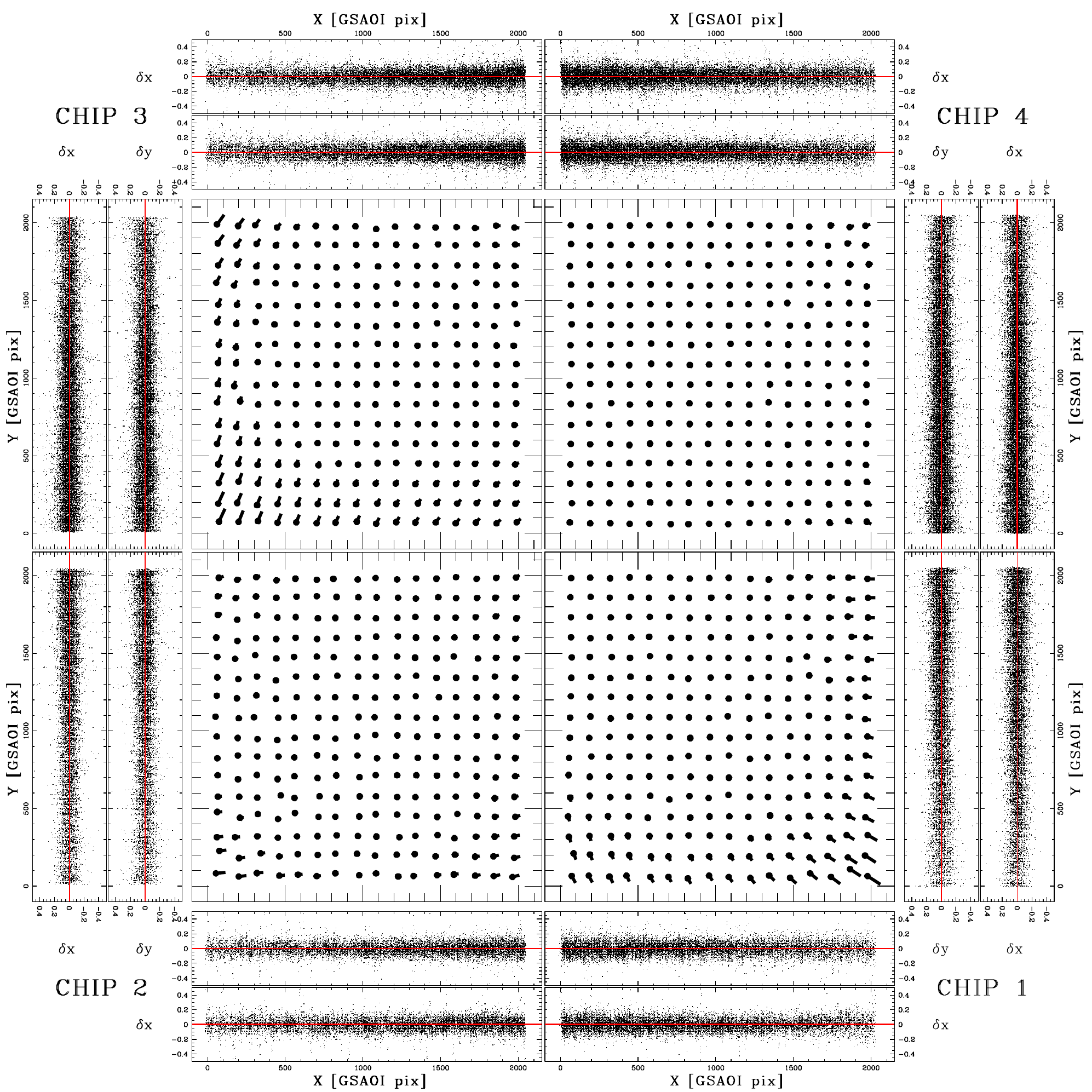}}
\caption{Residual map of the four chips of GSAOI camera, in the $K_s$ filter, after the GD correction. 
Residual vectors are magnified by a factor of 5000.}
\label{gdrk}
\end{figure*}
\begin{figure*}[!htbp]
\centering
{\includegraphics[width=17cm, angle=0]{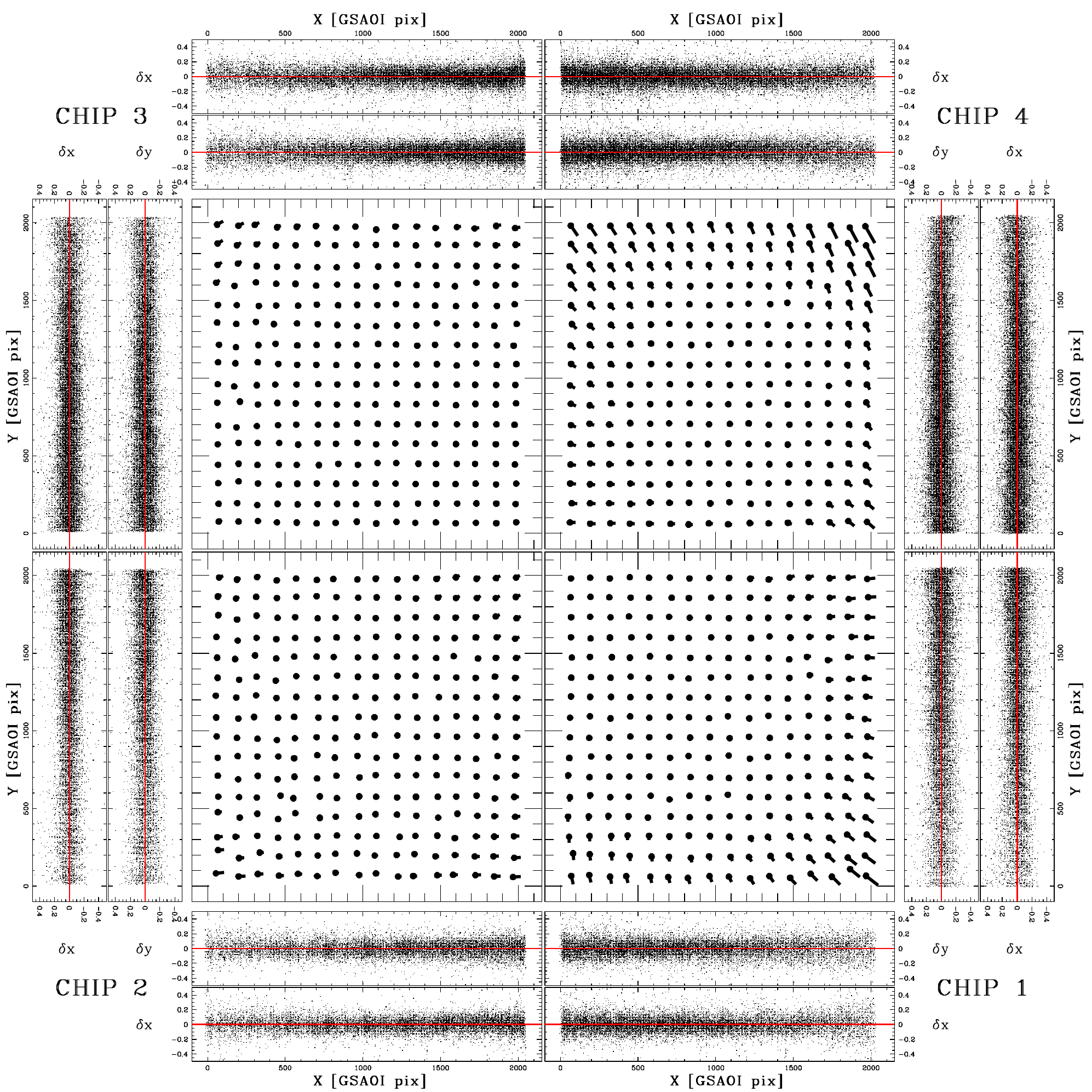}}
\caption{As in Figure~\ref{gdrk}, but for the $J$ band.}
\label{gdrj}
\end{figure*}
\end{document}